\documentclass[5p,twocolumn]{elsarticle}

\usepackage{graphicx}
\usepackage{dcolumn}
\usepackage{pifont}
\usepackage{bm}
\usepackage{multirow}
\usepackage{amsmath}
\usepackage{float}
\usepackage{txfonts}
\usepackage[version=3]{mhchem}

\usepackage[usenames,dvipsnames]{xcolor}
\definecolor{blue}{RGB}{0,0,225}
\definecolor{cream}{RGB}{222,217,201}
\definecolor{red}{RGB}{225,0,0}

\journal{arXiv}

\begin{document}
\title{First-principles study on material properties and stability of inorganic halide perovskite solid solutions CsPb(I$_{1-x}$Br$_x$)$_3$ towards high performance perovskite solar cells}

\author[kimuniv-m]{Chol-Jun Yu\corref{cor}}
\cortext[cor]{Corresponding author}
\ead{cj.yu@ryongnamsan.edu.kp}
\author[kimuniv-m]{Un-Hyok Ko}
\author[kimuniv-m]{Suk-Gyong Hwang}
\author[kimuniv-m]{Yun-Sim Kim}
\author[kimuniv-m]{Un-Gi Jong}
\author[kimuniv-m]{Yun-Hyok Kye}
\author[kimuniv-m]{Chol-Hyok Ri}

\address[kimuniv-m]{Chair of Computational Materials Design, Faculty of Materials Science, Kim Il Sung University, Ryongnam-Dong, Taesong District, Pyongyang, Democratic People's Republic of Korea}

\begin{abstract}
All-inorganic halide perovskites have attracted a great interest as a promising light harvester of perovskite solar cells due to their enhanced chemical stability.
In this work we investigate the material properties of solid solutions \ce{CsPb(I$_{1-x}$Br$_x$)3} in cubic phase by applying the virtual crystal approximation approach within a density functional theory framework.
First we check the validity of constructed pseudopotentials of the virtual atoms (X = I$_{1-x}$Br$_x$) by verifying that the lattice constants follow the linear function of mixing ratio.
We then suggest an idea of using the hybrid HSE functional with linear increasing value of exact exchange term as increasing the Br content $x$, which produces the band gaps of \ce{CsPbX3} in good agreement with the available experimental data.
The calculated light absorption coefficients and reflectivity show the systematic varying tendency to the Br content.
We calculate the phonon dispersions of \ce{CsPbX3}, CsX and \ce{PbX2} as slightly changing their volumes, revealing the phase instability of \ce{CsPbX3} and calculating the thermodynamic potential function differences.
By projecting Gibbs free energy differences onto the plane of $\Delta G=0$, we determine the $P-T$ diagram for \ce{CsPbX3} to be stable against the chemical decomposition, highlighting that the area of being stable extends gradually as the Br content increases.

\end{abstract}

\maketitle

\section{Introduction}
Recently, halide perovskites either in hybrid organic-inorganic or all-inorganic form have attracted a great interest in applications of photovoltaic perovskite solar cells (PSC)~\cite{Li18nrm, Saliba118s} and photoluminescent light emitting diodes (LED)~\cite{Davis17jpcc, Protesescu15nl}.
In spite of their excellent optical, optoelectronic properties and thus high efficiencies, however, the hybrid halide perovskites, represented by \ce{MAPbX3} and \ce{FAPbX3} (MA = \ce{CH3NH3}, FA = \ce{CH(NH2)2}, X = I, Br, Cl), have been shown to be unavoidable from the severe degradation of device performance upon exposure to moisture, light and heat, mostly being attributed to the organic moieties~\cite{Leijtens17jmca}.
In order to improve the stability of devices, therefore, many researchers turned their attention to the all-inorganic halide perovskites, including caesium halide perovskites \ce{CsMX3} (M = Pb, Sn).
Especially, the cubic phase \ce{CsPbI3} is likely to be the most promising candidate for light harvester of PSCs~\cite{Li18nc, Eperon15jmca, Niu17ra, Ahmad17sp, Chen17am, Sutton16aem} due to its proper band gap of 1.73 eV, absorbing most of the visible-light spectrum up to 700 nm~\cite{Bekenstein15jacs}, ultra-fast dynamics of charge carrier~\cite{Mondal17ns}, extra-long carrier diffusion length over 1.5 $\mu$m~\cite{Li18nc}, and lifetimes exceeding 10 $\mu$s~\cite{Hutter17ael}.
Eperon et al.~\cite{Eperon15jmca} firstly fabricated the working \ce{CsPbI3} solar cells with an efficiency of 2.9\% in 2015, and soon after, PSCs with enhanced efficiencies of 10.7\%~\cite{Li18nc, Swarnkar16s}, 11.8\%~\cite{Zhang17sa}, 13.4\%~\cite{Sanehira17sa} and 15\%~\cite{Wang18nc} and excellent thermal/moisture stability have been realized by applying the quantum dot and surface passivation engineering.

It is well-known that the perovskites undergo a series of phase transitions from cubic ($\alpha$) to tetragonal ($\beta$) and to orthorhombic ($\gamma$) phases upon cooling.
Likewise, the cubic $\alpha$-phase \ce{CsPbI3} perovskite (black phase) is stable at high temperature above 310 $^\circ$C~\cite{Trots08jpcs}, while at room temperature it exhibits another orthorhombic $\delta$-phase (yellow phase), known as a non-perovskite polymorph (perovskitoid) with a wide band gap unsuitable for solar cell application.
Through temperature-dependent synchrotron X-ray diffraction (SXRD), Marronnier and co-workers~\cite{Marronnier18an, Stoumpos15acs} found that upon heating the yellow $\delta$-phase transformed to the black $\alpha$-phase at 587 K, and subsequently upon cooling, it converted to the tetragonal $\beta$-phase at 554 K and the orthorhombic $\gamma$-phase at 457 K, which are also known as the black perovskite phases with some \ce{PbI6} octahedral tilting.
To realize the high efficient PSCs and LEDs using \ce{CsPbI3}, the cubic $\alpha$-phase should be stabilized at room temperature.
In doing so, it is important to keep in mind that the grain size of \ce{CsPbI3} crystal should be small like in forms of nanoplatelets and nanocrystals or nanocrystal quantum dots (QDs)~\cite{Gan19ael, Li17s}.
This can be accomplished by carefully controlling the synthesis process~\cite{Eperon15jmca, Hutter17ael, Swarnkar16s, Zhang17sa, Dayan19cm, Straus19jacs} and by using effective surface capping ligands like oleic acid and oleylamine~\cite{Protesescu15nl, Long17cc}, alkyl phosphinic acid~\cite{Wang17cc}, and poly-vinylpyrrolidone~\cite{Li18nc}.
With these techniques, the $\alpha$-\ce{CsPbI3} thin films can be stabilized at room temperature, far below the phase transition temperature, and function under ambient condition for several months.
However, these synthesizing methods require heating and inert gas protection using vacuum chamber and glove box, resulting in an increase of fabrication cost.

Halide anion exchange has proven to be a very promising way for producing the cubic \ce{CsPbI3} nanocrystals and even bulk at low temperature, taking advantage of high mobility of halide anions~\cite{Nedelcu15nl, Akkerman15jacs, Hoffman16jacs, Zhang17cec, Li16ra}.
To do this, the cubic \ce{CsPbBr3} nanocrystals were first prepared and then soaked in a \ce{PbI2} or \ce{PbCl2} solution, followed by a fast, partial or complete anion exchange of Br-to-I or Br-to-Cl at low temperature of 40 $-$ 75 $^\circ$C without change of cubic phase and crystal shape~\cite{Nedelcu15nl, Akkerman15jacs}.
The employment of zinc halogenide (\ce{ZnX2}) instead of \ce{PbX2} was reported to reduce the reaction temperature to room temperature with remarkably short reaction time~\cite{Zhang17cec}.
In particular, Kamat et al.~\cite{Hoffman16jacs} developed an alternative layer-by-layer deposition procedure for fabricating $\alpha$-\ce{CsPbBr3} bulk by sintering nanocrystals, which subsequently underwent the Br-to-I exchange at a solution temperature of 75 $^\circ$C to form the cubic \ce{CsPbI3} thin film with a thickness of 75 nm.
The activation energy was estimated to be 0.46 $\pm$ 0.06 eV for expanding the lattice of \ce{CsPbBr3} to incorporate I anions~\cite{Hoffman16jacs}, and the high speed of anion exchange was discussed to be rooted in the ionic properties of metal halide perovskites and fast diffusion of halide vacancies with the activation energy of 0.25 eV for \ce{CsPbBr3}~\cite{Nedelcu15nl}.
It should be noted that during the anion exchange their solid solutions \ce{CsPb(Br_xI_{1-x})3} are not excluded to form, especially in thicker film.
Rather, to avoid the phase degradation of \ce{CsPbI3} under ambient conditions, many researchers preferred to develop bromide-based PSCs such as \ce{CsPbBr3} (efficiency, 8.8\%)~\cite{Kulbak15jpcl, Qian18apl, Liu19ne}, \ce{CsPbIBr2} (9.8\%)~\cite{YGuo19jmca, ZGuo19jmca, Zhu18aem, Lau16ael}, \ce{CsPbI2Br} (16.1\%)~\cite{Chen19j} and \ce{CsPb(Br_xI_{1-x})3} (x = 0.66, 14.1\%)~\cite{Parida19jmca, Ng18sr, Zhao19jpcc, Li19jpcl}.
In spite of such intensive experimental studies, theoretical works are yet absent and thus a persuasive explanation for halide anion exchange is not provided.

As such, modelling and simulations based on density functional theory (DFT) have proven to be a powerful tool for designing perovskite materials~\cite{yu19jpe}.
We have reported several works for all-inorganic halide solid solutions \ce{(Rb$_x$Cs$_{1-x}$)PbI3}~\cite{Jong18prb, Jong18jmca2} as well as hybrid cousins \ce{MAPb(I$_{1-x}$Br$_x$)3}~\cite{Jong16prb} and \ce{MAPb(I$_{1-x}$Cl$_x$)3}~\cite{Jong17jps} by applying virtual crystal approximation (VCA) approach.
With DFT calculations, electronic structures and optical properties of \ce{CsMX3} have been investigated by several authors~\cite{Brgoch14jpcc,Ying18jpcc}, and Walsh et al.~\cite{RYang17jpcl} identified octahedral tilting in the cubic \ce{CsPbX3} perovskites.
Through defect calculations considering solvation effect, we revealed the stabilization of $\alpha$-\ce{CsPbI3} driven by vacancy~\cite{Kye19jpcc}.
The lattice vibrational properties of \ce{CsPbI3} polymorphs were investigated by calculating phonon dispersions within density functional perturbation theory (DFPT), identifying anharmonic phonon modes in the cubic phase~\cite{Marronnier18an, Marronnier17jpcl}.
In this work, we aim to clarify variation tendency of structural, electronic, optical and vibrational properties of solid solutions \ce{CsPbX3} in cubic phase as varying the Br content $x$ with DFT calculations.

\section{Methods}
For the DFT calculations in this work, we applied the pseudopotential plane-wave method as implemented in Quantum ESPRESSO (QE, version 6.2.0)~\cite{QE} and ABINIT (version 8.8.4)~\cite{abinit16,abinit09} packages.
Troullier-Martins type soft norm-conserving pseudopotentials of all the relevant atoms~\cite{TMpseudo} were constructed using the input files provided in the pslibrary (version 1.0.0) and LD1 code included in the QE package~\cite{QE}. 
Valence electron configurations of the atoms are Cs-$6s^16p^0$, Pb-$4f^{14}5d^{10}6s^26p^2$, I-$4d^{10}5s^25p^5$ and Br-$3d^{10}4s^24p^5$.
The Perdew-Zunger (PZ) formalism~\cite{pz81prb} within the local density approximation (LDA) was used to express the exchange-correlation (XC) interaction, and scalar relativistic effect was considered for all the atoms.
The pseudopotentials of virtual atoms $X$ = I$_{1-x}$Br$_x$ ($x=0.0,~0.1,~...~,~1.0$) were constructed by using the modified virtual.x code in the QE package, according to the $YE^2A^2$ formalism~\cite{yucj07jpcsm}.

For structural optimizations and phonon calculations, we used the QE package with reasonable computational parameters; kinetic cutoff energies of 60 Ry and 480 Ry for the plane wave basis sets and electron charge density, and special $k$-points of ($6\times6\times6$) and ($6\times6\times4$) for cubic and hexagonal unit cells, respectively.
These computational parameters guarantee accuracies of 0.5 meV per formula unit for the total energy and 0.1 meV for the phonon energy, respectively.
We calculated the DFT total energies ($E$) of the unit cells with evenly increasing volumes from 0.9$V_0$ to 1.1$V_0$ with 11 intervals, where $V_0$ is the equilibrium volume, and fitted the resultant $E-V$ data to the 4th-order natural strain equation of state (EOS) for crystalline solid~\cite{Poirier98pepi}.
This procedure provided the equilibrium lattice-related properties such as optimal lattice constants and bulk modulus, and the pressure curve versus volume necessary for enthalpy calculation.
For the case of hexagonal lattice, the atomic positions were allowed to relax until the atomic forces converged to 0.01 eV/\AA.
Using the optimized unit cells, we calculated phonon dispersions of the relevant crystals with the different volumes at each Br content $x$, obtaining the entropy and free energy ($F$) within the quasiharmonic approximation (QHA), and fitted the $F-V$ data to EOS, resulting in the Gibbs free energy and temperature dependent lattice constant and bulk modulus, as described in our previous works~\cite{Jong18jmca2,Yu14pb}.
Based on the calculated Gibbs free energies, we estimated the miscibility of solid solutions, defined by the free energy difference,
\begin{equation}
\Delta G=G_{\ce{CsPbX3}}-(1-x)G_{\ce{CsPbI3}}-xG_{\ce{CsPbBr3}}
\label{eq-miscib}
\end{equation}
where $X$ = I$_{1-x}$Br$_x$.
The negative values indicate that the solid solutions \ce{CsPbX3} are more favorable thermodynamically than their individual components.
Meanwhile, the thermodynamic stability of \ce{CsPbX3} in opposition to their decomposition into \ce{CsX} and \ce{PbX2} at finite temperature and pressure was predicted by the Gibbs free energy difference as follows,
\begin{equation}
\Delta G=G_{\ce{CsPbX3}}-G_{\ce{CsX}}-G_{\ce{PbX2}}
\label{eq-gibbs}
\end{equation}
where \ce{CsPbX3} and CsX were assumed to be in the cubic phase with a space group of $Pm\bar{3}m$ while \ce{PbX2} in the hexagonal phase with a space group of $P\bar{3}m1$.
The Perdew-Burke-Ernzerhof functional for solids (PBEsol)~\cite{PBEsol} within the generalized gradient approximation (GGA) was used for the XC interaction.

To calculate the electronic band structures and optical properties of the solid solutions \ce{CsPbX3}, we used the ABINIT package with a cutoff energy of 80 Ry and ($6\times6\times6$) $k$-points.
We also applied the Heyd-Scuseria-Ernzerhof (HSE) hybrid functional~\cite{HS04jcp} with gradually increasing portion of exact Hartree-Fock (HF) exchange functional from 0.0 to 0.33 as increasing the Br content $x$ in order to obtain electronic band structures and band gaps in good agreement with experiment.
The frequency dependent dielectric constants, $\varepsilon(\omega)=\varepsilon_1(\omega)+i\varepsilon_2(\omega)$, were calculated by solving the Bethe-Salpeter equation with an excitonic effect (BSC-EXC) as implemented in the ABINIT package.
For comparison, we also performed the density functional perturbation theory (DFPT) calculations within the random phase approximation in the Kohn-Sham approach (RPA-KS) and the $GW$ approach (RPA-GW).
Then, absorption coefficient as a function of photon energy $\alpha(\omega)$ was given by the following equation~\cite{Jong18prb,Jong16prb,Jong17jps},
\begin{equation}
\alpha(\omega)=\frac{\sqrt{2}\omega}{c}\left[\sqrt{\varepsilon_1^2(\omega)+\varepsilon_2^2(\omega)}-\varepsilon_1(\omega)\right]^{\frac{1}{2}}
\label{eq-absorp}
\end{equation}
where $c$ is the light velocity, and reflectivity $R(\omega)$ by the following equation~\cite{Ying18jpcc}:
\begin{equation}
R(\omega)=\left|\frac{\sqrt{\varepsilon_1(\omega)+i\varepsilon_2(\omega)}-1}{\sqrt{\varepsilon_1(\omega)+i\varepsilon_2(\omega)}+1}\right|^2
\label{eq-reflect}
\end{equation}
Due to computational limits, the scissor correction was obtained from PBEsol calculations and added to the final $GW$ calculations to take account of spin-orbit coupling (SOC) effect.

\section{Results and discussion}
\subsection{Assessing pseudopotentials of virtual atoms}
It has been accepted that the greatest merit of using the VCA approach in the research of solid solutions, when compared to its counterpart super cell approach, is to use the unit cell comprising the minimal number of atoms.
Therefore, if the pseudopotentials of virtual atoms once become proved to be reliable, material properties of solid solutions at any mixing ratio could be predicted with the lowest computational load.
The first check of pseudopotentials should be a verification of the Vegard's law, which indicates a linear dependence of lattice constants as increasing the mixing ratio.

\begin{figure}[!th]
\centering
\includegraphics[clip=true,scale=0.5]{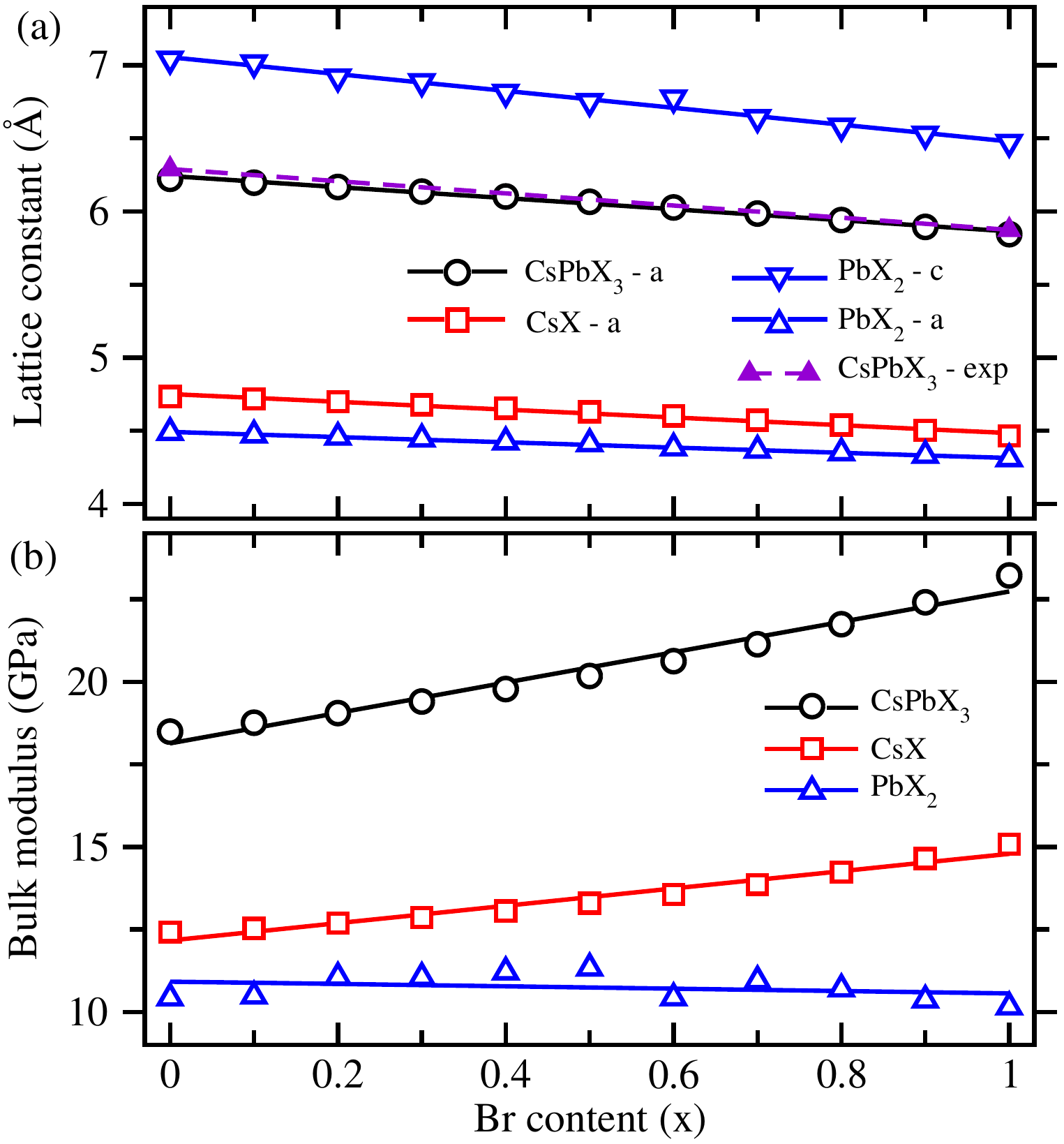}
\caption{(a) Lattice constant and (b) bulk modulus of cubic $\alpha$-\ce{CsPbX3}, cubic CsX and hexagonal \ce{PbX2} where X = \ce{I_{1-x}Br_x} as varying the Br content $x$, calculated by VCA approach.}
\label{fig1}
\end{figure}
In Fig.~\ref{fig1}, we show the lattice constants and bulk moduli of solid solutions \ce{CsPbX3}, CsX and \ce{PbX2} (X = \ce{I_{1-x}Br_x}) as functions of Br content $x$, obtained by using the pseudopotentials constructed in this work and fitting $E-V$ data to the 4th-order natural strain EOS.
For cubic \ce{CsPbX3}, the optimized lattice constants $a$ were found to follow the linear function of Br content such as $a(x)=6.242-0.378x$ (\AA), indicating a satisfaction of Vegard's law.
The calculated lattice constants of 6.297 \AA~for \ce{CsPbI3} and 5.874 \AA~for \ce{CsPbBr3} were confirmed to agree well with the experimental values with relative errors of $-0.87$\%~\cite{Marronnier18an} and $-0.17$\%~\cite{Akkerman15jacs}, respectively.
Similar tendencies were observed for the lattice constants $a$ for cubic CsX like $a(x)=4.751-0.267x$ (\AA), and $a$ and $c$ for hexagonal \ce{PbX2} like $a(x)=4.493-0.179x$ (\AA) and $c(x)=7.055-0.576x$ (\AA).
As shown in Fig.~\ref{fig1}(b), the bulk moduli were found to increase for cubic phases but slightly decrease for hexagonal phase as increasing the Br mixing ratio.
It should be noted that such shrinking of cubic lattice and increase of bulk modulus when mixing bromide with iodide perovskite could result in strengthening of Pb-X bond and thus thermodynamic stability as will be shown below.
Based on the positive consequence for assessing the pseudopotentials of virtual atoms, we can safely progress the work to find the variation trend of material properties of solid solutions with the VCA approach.

\subsection{Electronic and optical properties}
\begin{figure*}[!th]
\centering
\begin{tabular}{cc}
\hspace{4pt}\includegraphics[clip=true,scale=0.5]{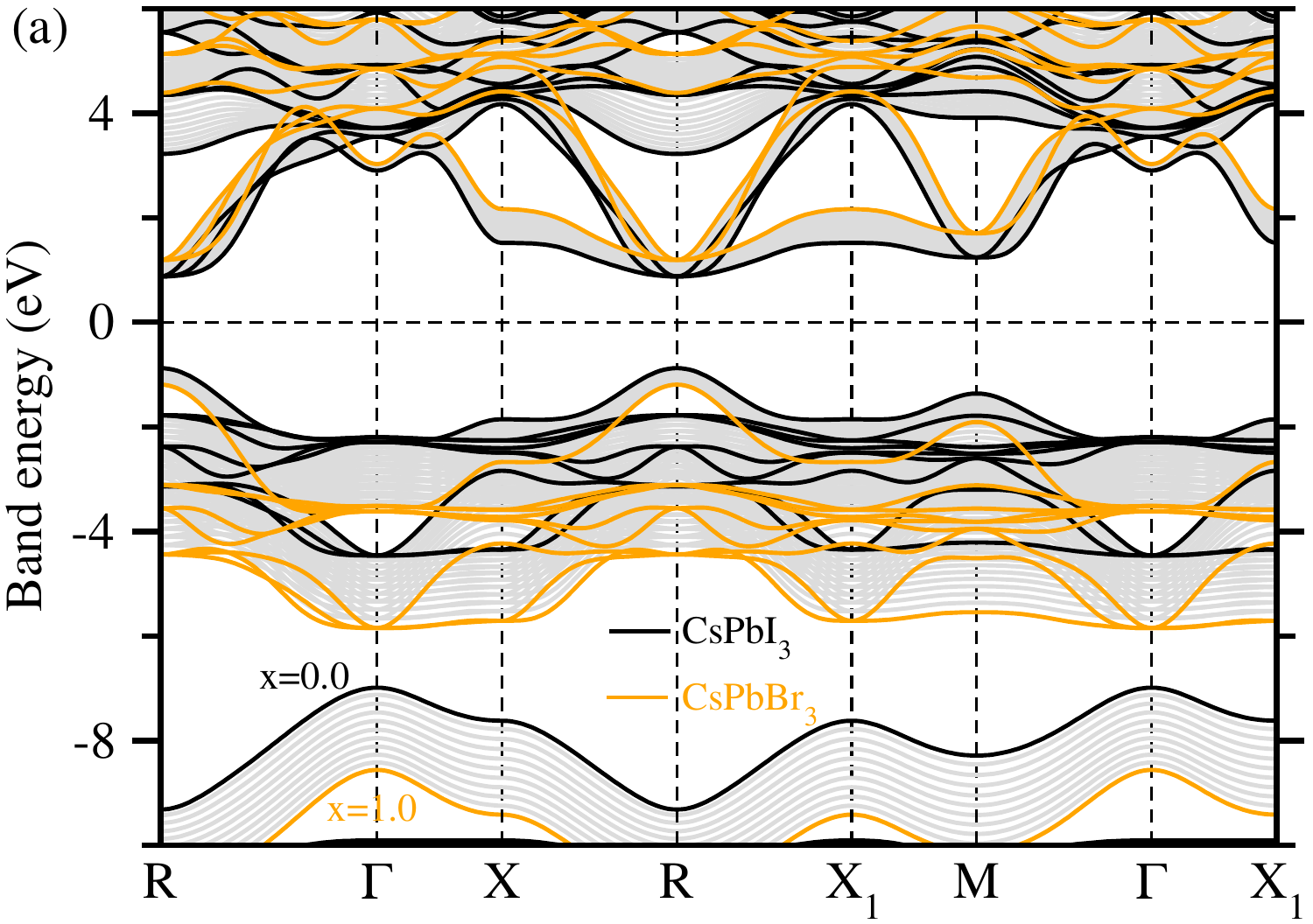} &
\hspace{21pt}\includegraphics[clip=true,scale=0.075]{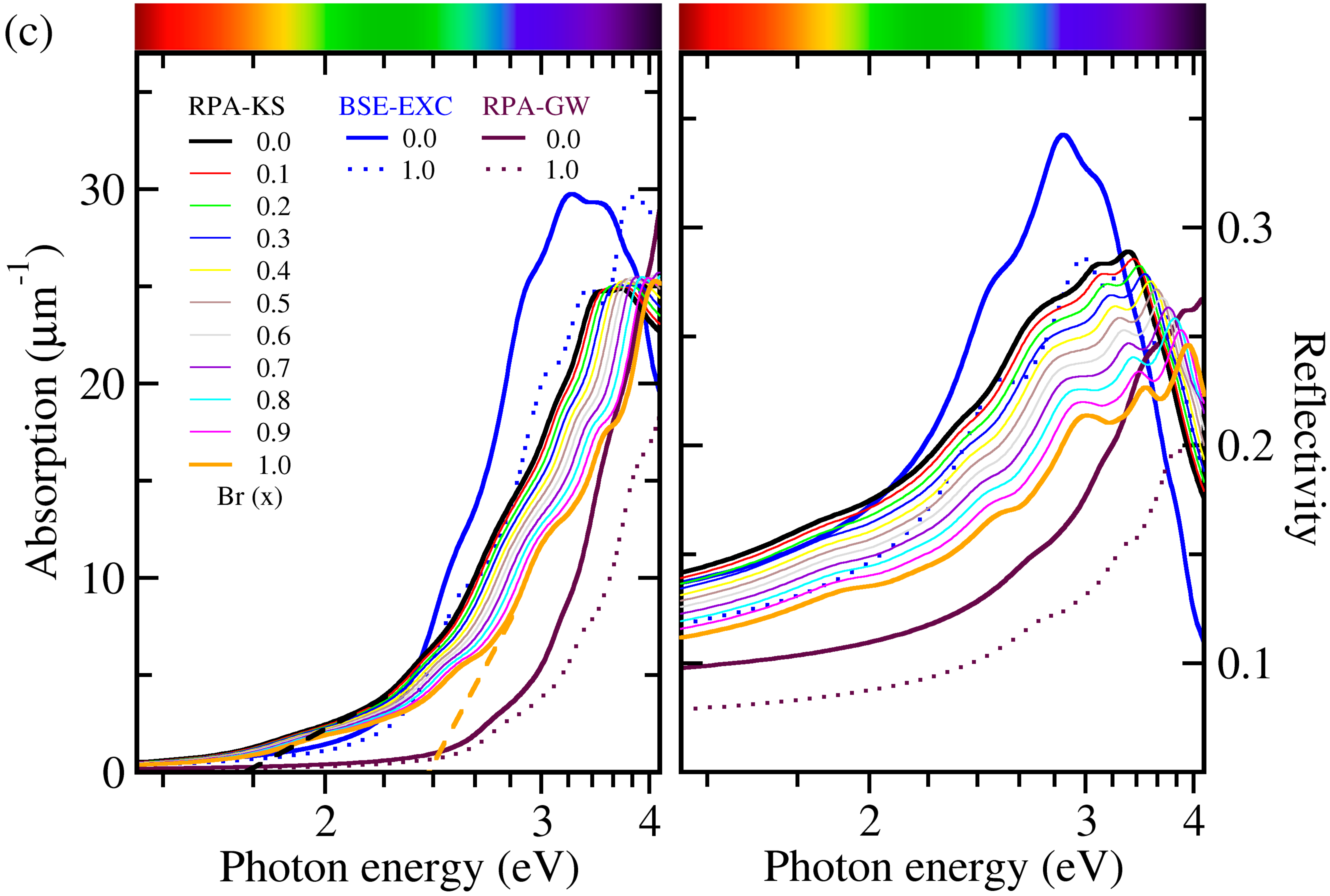} \\
\includegraphics[clip=true,scale=0.5]{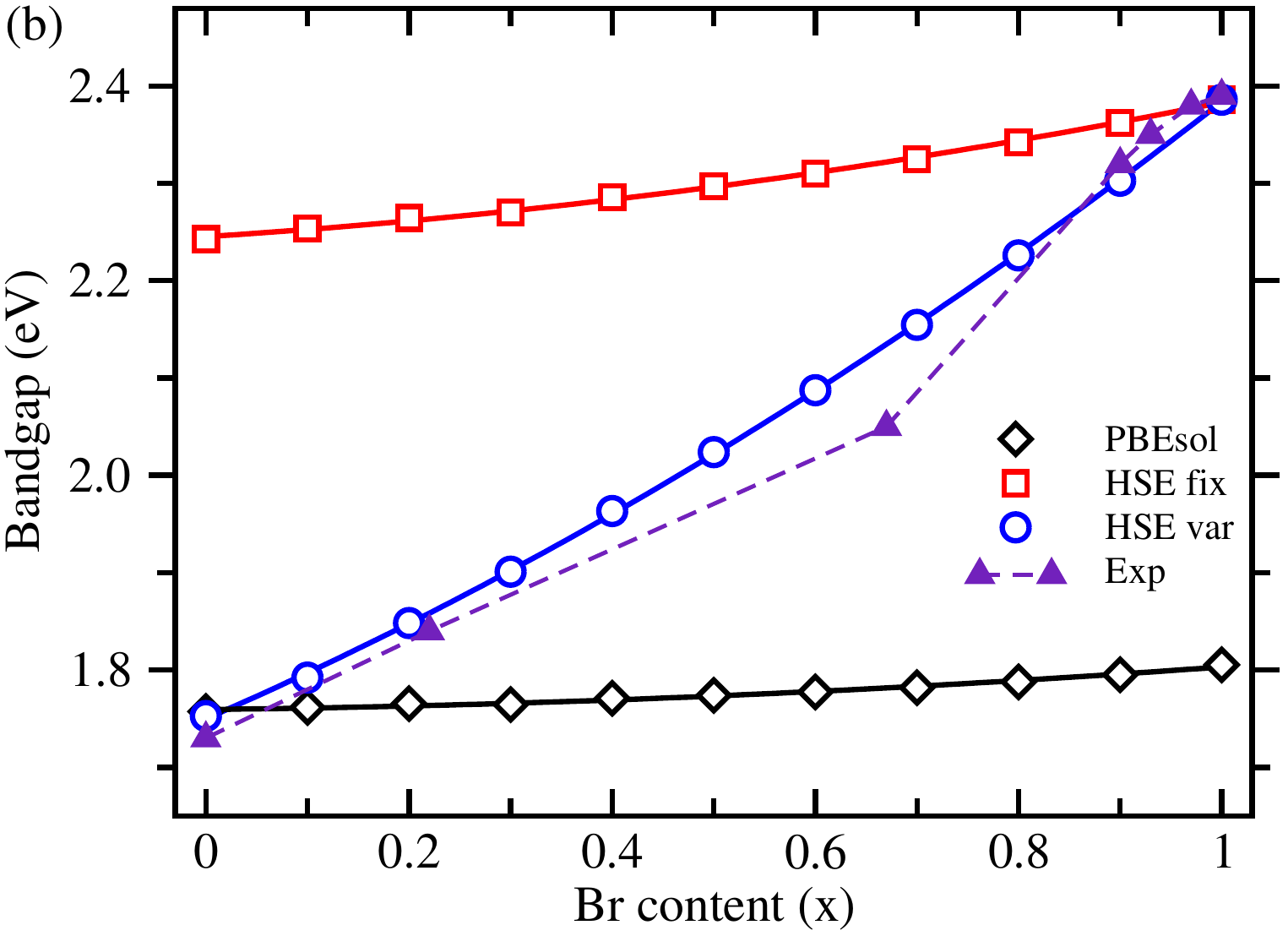} &
\includegraphics[clip=true,scale=0.5]{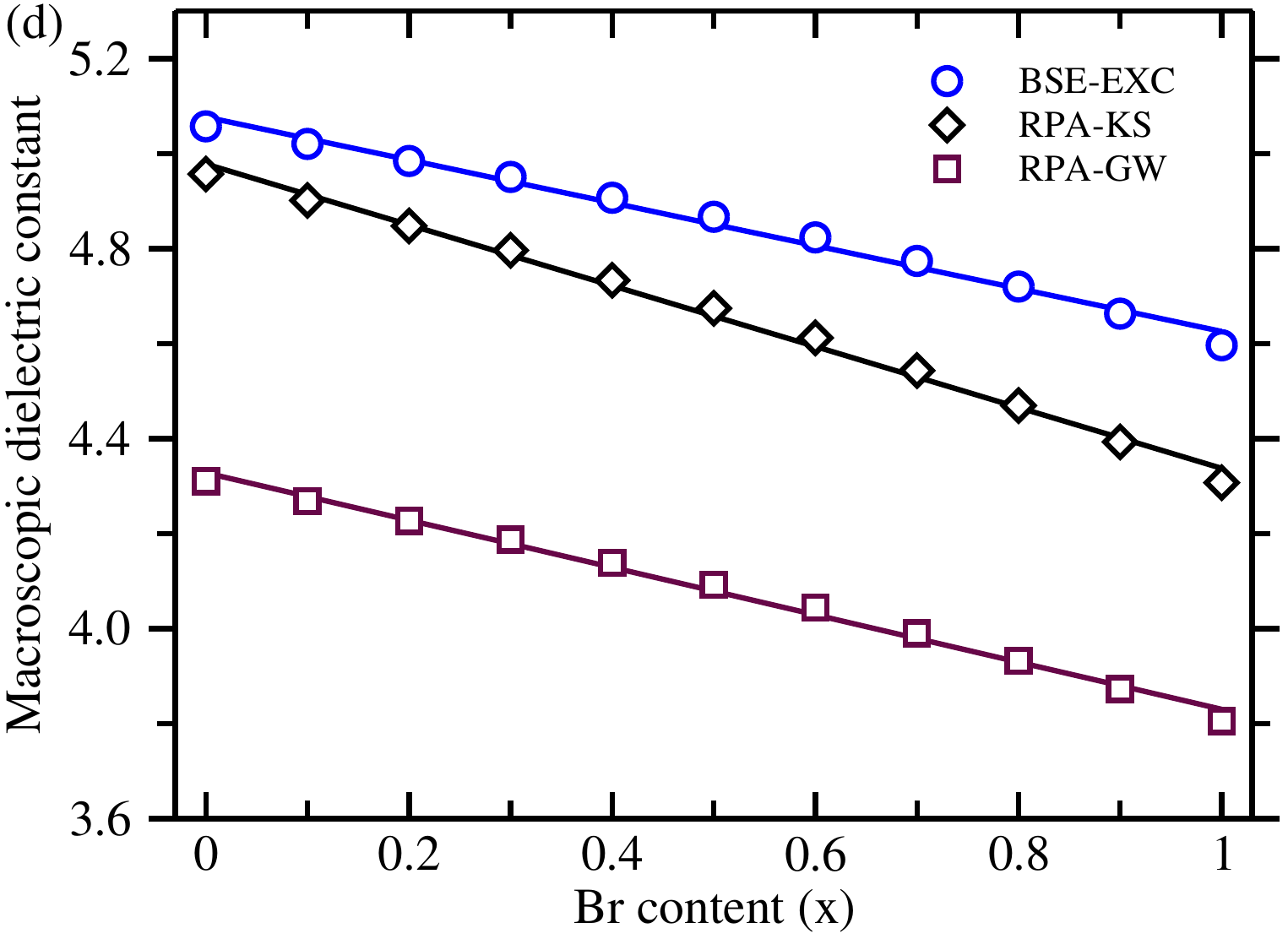}
\end{tabular}
\caption{(a) Electronic band structures of cubic \ce{CsPbX3} (X = I$_{1-x}$Br$_x$), calculated by using the HSE hybrid functional with linearly increasing Hartree-Fock (HF) exchange portion from 0.01 to 0.33, and (b) their band gaps as a function of Br content $x$, calculated by using PBEsol, HSE with fixed HF portion of 0.33 and HSE with linear increase of HF portion. Experimental values are from ref.~\cite{Wang18nc} for \ce{CsPbI3},  ref.~\cite{Parida19jmca} for CsPbI$_{3-x}$Br$_x$ ($x=0.66$), ref.~\cite{Zhu18aem, Lau16ael} for \ce{CsPbIBr2} and ref.~\cite{Liu19ne} for \ce{CsPbBr3}. (c) Light absorption coefficient (left panel) and reflectivity (right panel) as a function of photon energy, and (d) macroscopic dielectric constant as a function of Br content $x$, calculated by solving Bethe-Salpeter equation (BSE) considering excitonic (EXC) effect, GW and Kohn-Sham (KS) equations within random phase approximation (RPA).}
\label{fig2}
\end{figure*}

As for electronic properties, we calculated the band structures of \ce{CsPbX3} solid solutions and their band gaps, which are critical factors in determining whether the material under study would be a promising light absorber for solar cell applications.
Since the band gaps of semiconductors are strongly dependent on the choice of XC functional in DFT calculations, we need to find the proper XC functional or approach that can reproduce the experimental band gaps.
Through many previous DFT calculations of halide perovskites, it has become clear that the PBEsol or PBE functional can give the band gaps of lead iodide perovskites such as MAPI and \ce{CsPbI3} in good agreement with experiment~\cite{Jong18prb,Jong16prb,Jong17jps} thanks to a fortuitous compensation of GGA underestimation and overestimation by the lack of SOC effect, which is in particular important for heavy elements Pb and I.
When including the SOC effect in PBE-GGA and HSE hybrid functionals, the band gaps were found to be quite underestimated by about 1 eV for \ce{CsPbI3} due to a split of Pb $6p$ states and their down-shift by the SOC effect~\cite{Jong18prb,Brgoch14jpcc}.
Therefore, we excluded the SOC effect from further consideration, and obtained the band structures of the two end compounds \ce{CsPbI3} and \ce{CsPbBr3}, using both PBEsol-GGA functional and HSE hybrid functional with exact HF exchange portion of 0.33 (see Fig. S3).
In accordance with the previous works for cubic \ce{CsPbI3}, the PBEsol functional provided a band gap of 1.76 eV in excellent agreement with experimental value of 1.73 eV~\cite{Eperon15jmca}, but the HSE functional yielded an overestimating value of 2.24 eV due to an up-shift of conduction bands and down-shift of valence bands.
For cubic \ce{CsPbBr3}, on the contrary, HSE gave a reliable band gap of 2.39 eV compared to the experimental value of 2.36 eV~\cite{Kulbak16jpcl}, while PBEsol yielded an underestimating value of 1.81 eV.
Such underestimation can be attributed to lighter element Br and thus weaker SOC effect than I.

\begin{figure*}[!th]
\centering
\begin{tabular}{cc}
\includegraphics[clip=true,scale=0.5]{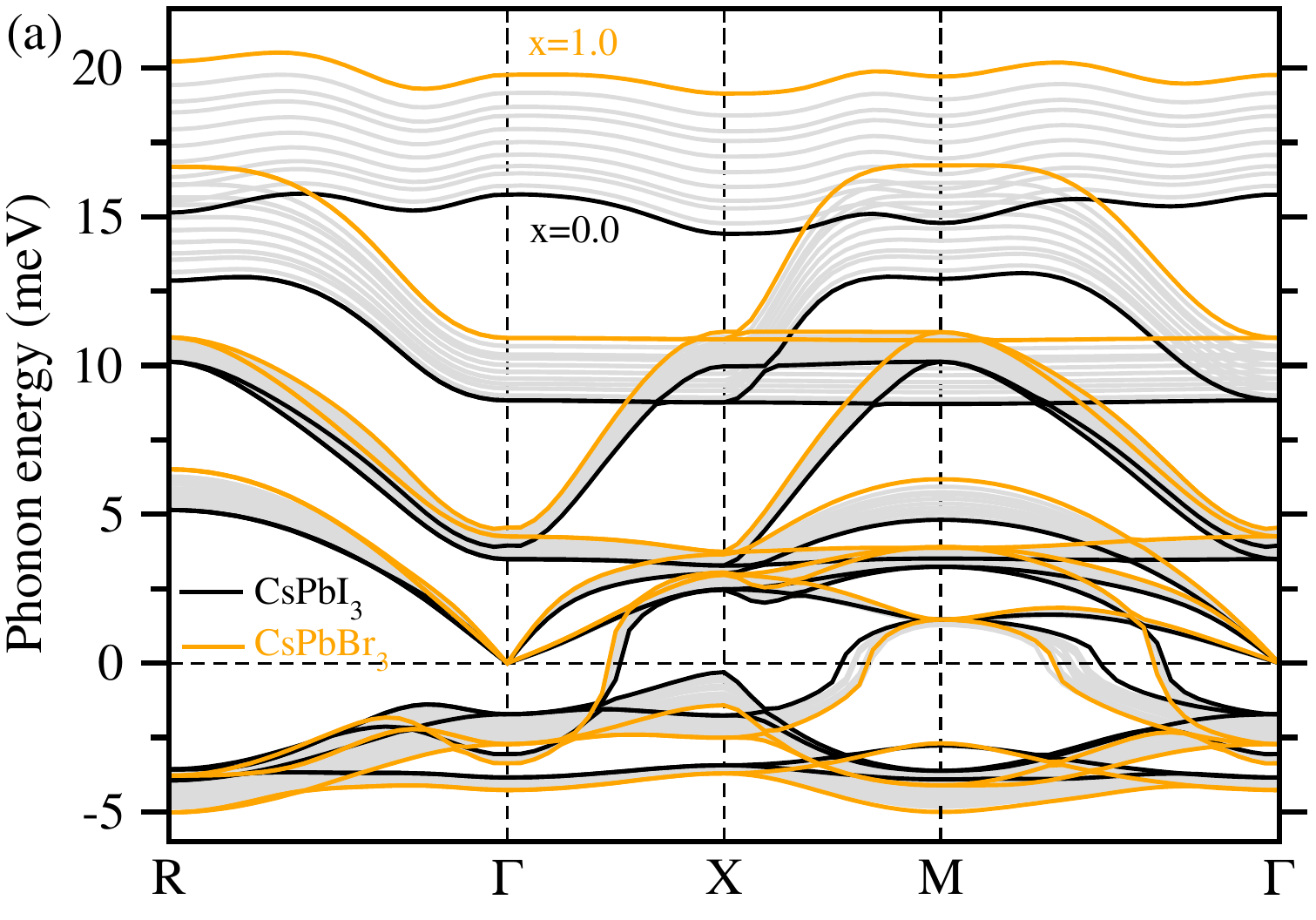} &
~~~\includegraphics[clip=true,scale=0.5]{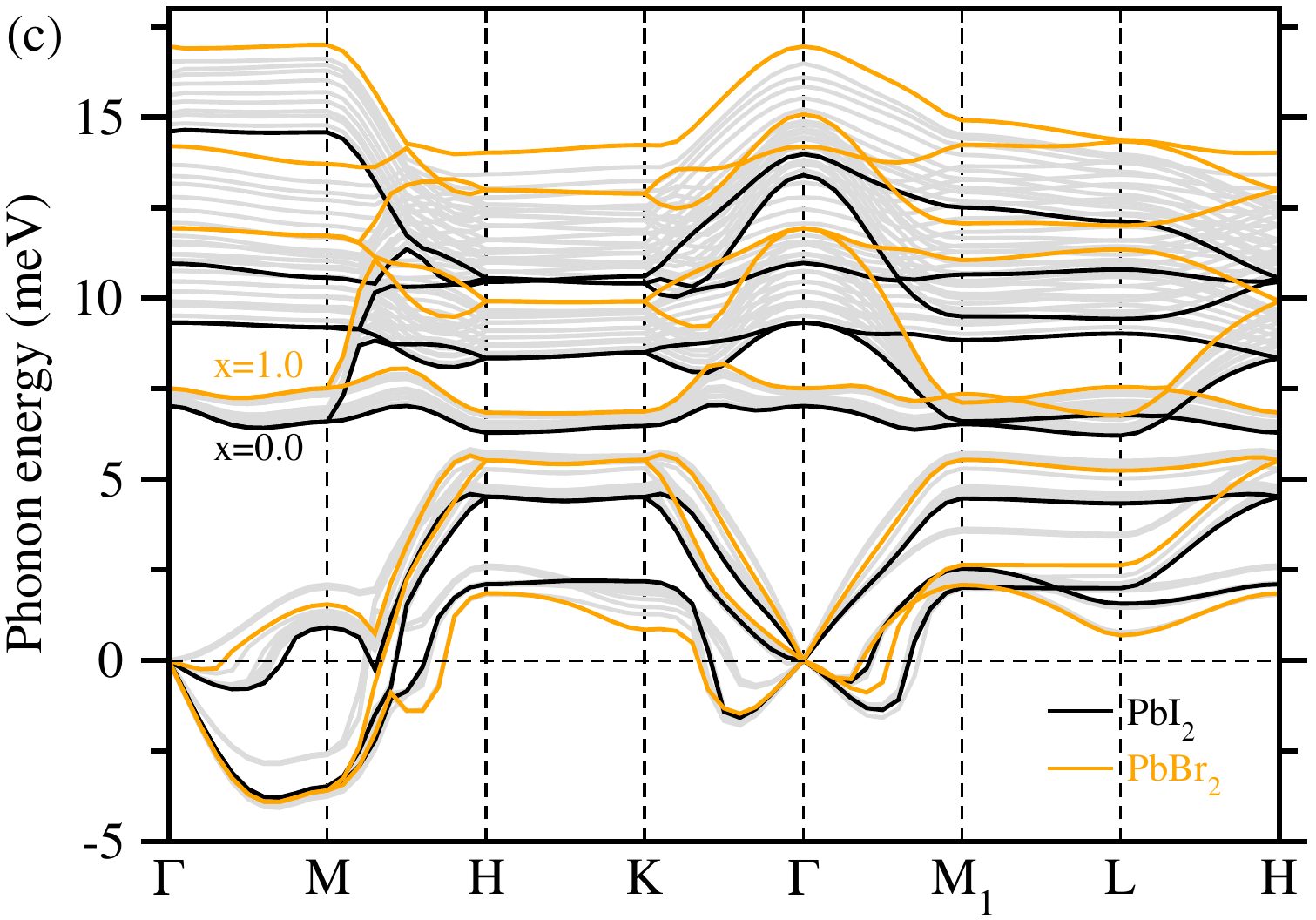} \\
\includegraphics[clip=true,scale=0.5]{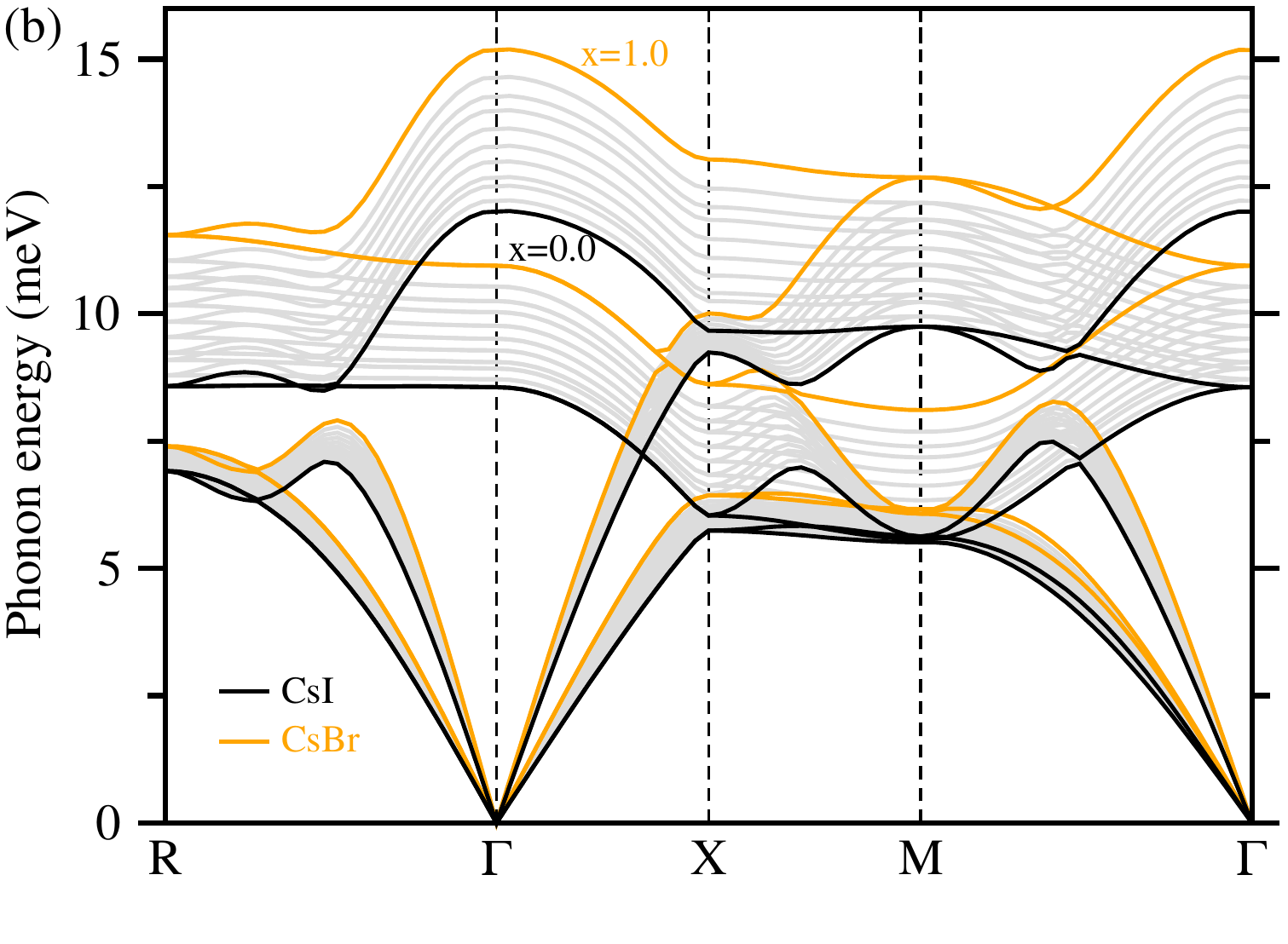} &
\includegraphics[clip=true,scale=0.5]{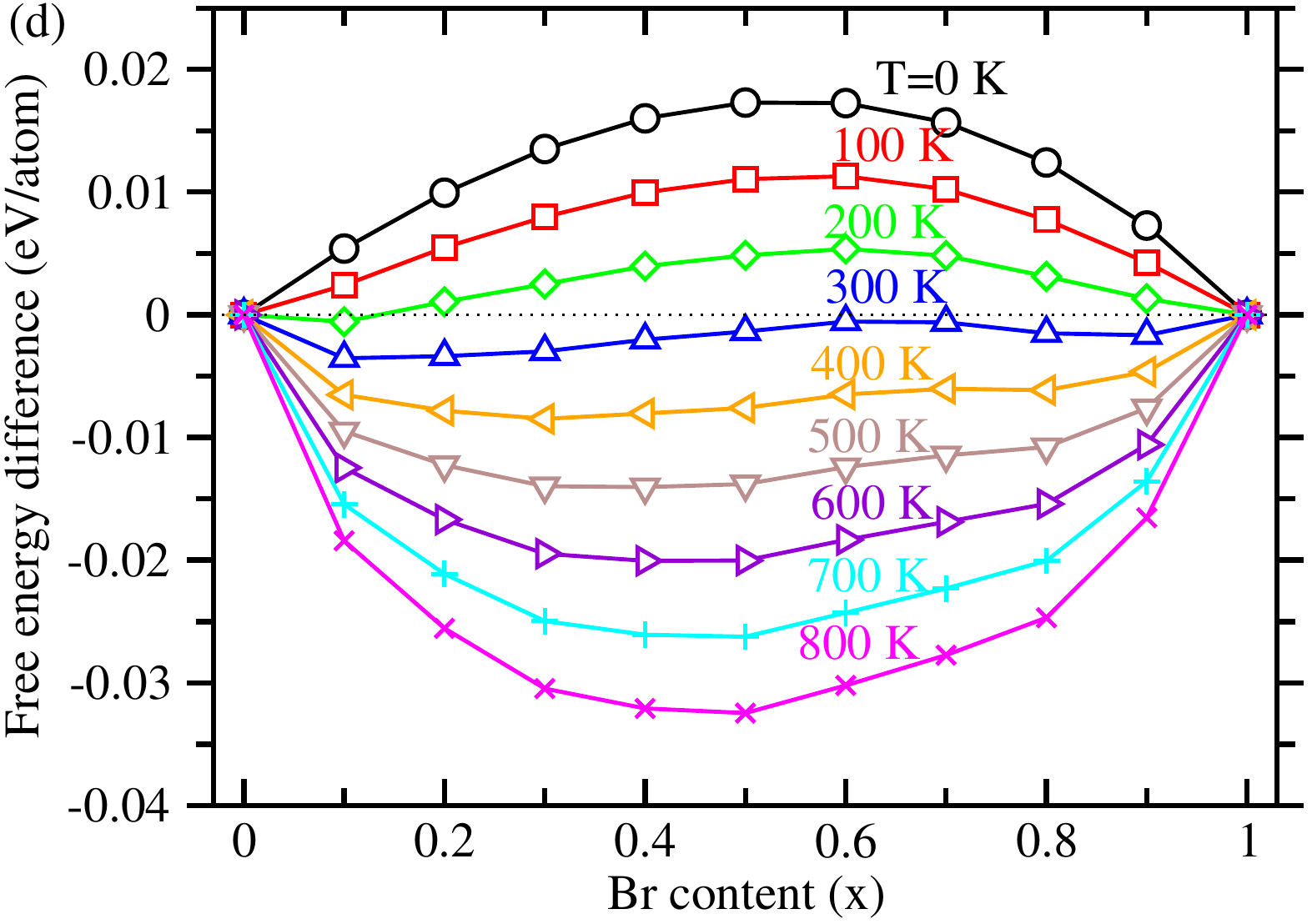}
\end{tabular}
\caption{Phonon dispersion curves of (a) cubic \ce{CsPbX3}, (b) cubic \ce{CsX} and (c) hexagonal \ce{PbX2} (X = I$_{1-x}$Br$_x$) as increasing the Br content $x$ from 0.0 to 1.0, calculated by DFPT approach. (d) Gibbs free energy of mixing \ce{CsPbBr3} and \ce{CsPbI3} to form their solid solutions \ce{CsPbX3} at various temperatures.}
\label{fig3}
\end{figure*}

In order to obtain reasonable band gaps of solid solutions \ce{CsPbX3}, we decided to calculate their electronic band structures by using the HSE hybrid functional with a linearly increasing HF exchange portion from 0.01 to 0.33 as increasing the Br content from $x=0$ to 1.
As shown in Fig.~\ref{fig2}(a), the conduction bands were observed to be gradually up-shifted whereas the valence bands were found to be systematically down-shifted as going from \ce{CsPbI3} to \ce{CsPbBr3}, leading to a gradual increase of band gaps as increasing the Br content in \ce{CsPbX3} solid solutions.
Figure~\ref{fig2}(b) presents the band gaps of solid solutions as a function of Br content $x$ calculated with different XC functionals.
It was found that the PBEsol functional gave much lower band gaps for higher Br contents whereas the HSE hybrid functional with fixed exchange portion of 0.33 yielded much higher band gaps for lower Br contents.
When using the HSE functional with the linear increasing exchange portions from 0.01 to 0.33, the band gaps were calculated to be in good agreement with the available experimental data in overall Br contents.
Since the band gaps are likely to vary quadratically, the data was interpolated into a quadratic function of Br content, resulting in $E_g(x)=1.750+0.454x+0.180x^2$ (eV).
This gives the fitting coefficients such as $E_g(0)=1.750$ eV for \ce{CsPbI3}, $E_g(1)=2.384$ eV for \ce{CsPbBr3}, and the bowing parameter of 0.180 eV~\cite{Jong16prb,Jong17jps,Atourki16ass,Noh13nl}, which is comparable with 0.185 eV of the organic-inorganic hybrid counterpart MAPb(I$_{1-x}$Br$_x$)$_3$~\cite{Jong16prb} but much lower than 0.873 eV of MAPb(I$_{1-x}$Cl$_x$)$_3$~\cite{Jong17jps}.
Considering that the bowing parameter represents the nonlinear effect coming from the anisotropy of chemical binding and fluctuation degree in the crystal field, its smaller value indicates a lower compositional disorder and a good miscibility between \ce{CsPbI3} and \ce{CsPbBr3}.
It should be noted that the increase of band gap when replacing I with Br implies a detriment of the light harvesting properties and a possibly decline of PSC efficiency.

As for optical properties we calculated the frequency-dependent dielectric constants of \ce{CsPbX3} solid solutions by using the different levels of theory such as BSE-EXC, RPA-KS and RPA-GW (see Fig. S4 for real and imaginary parts).
Then, using Eqs.~\ref{eq-absorp} and~\ref{eq-reflect}, we determined their light absorption coefficients and reflectivity as increasing the Br content, as shown in Fig.~\ref{fig2}(c) for the relevant range of photon energy and Fig. S5 for the full range.
It was found that the RPA-KS method could produce the most reasonable absorption onsets, marked with dashed lines in Fig.~\ref{fig2}(c), for all the considered Br contents when compared with experiment~\cite{Bekenstein15jacs}, while the BSE-EXC and RPA-GW methods yielded forward- and backward-shifted curves in the axis of photon energy.
With these different methods with one accord, the first peaks of the absorption coefficients were found to be shifted backward and the reflectivity to descend gradually, indicating a slight enhancement of light absorption as increasing the Br content.
The static dielectric constants $\varepsilon_s$ were readily obtained from the real parts of the frequency-dependent dielectric constants by extracting data at the zero photon energy.
As shown in Fig.~\ref{fig2}(d), they decrease according to a linear function of Br content as $\varepsilon_s(x)=5.077-0.451x$ by BSE-EXC, $4.978-0.640x$ by RPA-KS and $4.328-0.498x$ by RPA-GW respectively, which are comparable with the organic-inorganic hybrid halide perovskites~\cite{Jong16prb,Jong17jps,Even14jpcc}.
Considering that the exciton binding energy is in inverse proportion to a square of the static dielectric constant~\cite{Jong16prb,Jong17jps}, the rate of recombination of photo-excited electron and hole can be enhanced as increasing the Br content in \ce{CsPbX3} solid solutions.

\subsection{Lattice vibration-related properties}
The major issue in perovskite solar cells is a device stability, which can be represented mostly by the material stability of halide perovskites upon exposure to moisture, heat and light.
To get an insight of material stability at ambient condition, we need to investigate the lattice vibrational properties through calculation of the phonon dispersions, which will be used in determination of atomic vibrational contributions to the free energy.
Figures~\ref{fig3}(a)-(c) show the phonon dispersion curves of solid solutions \ce{CsPbX3}, CsX and \ce{PbX2} along the high symmetric points of the Brillouin zone (BZ).
It should be noted that the splitting between longitudinal optic (LO) and transverse optic (TO) modes (LO-TO splitting) at the BZ centre $\Gamma$ point should be considered by taking into account the coupling between the ionic displacements and the homogeneous electric field generated in the polar insulators.

For the solid solutions \ce{CsPbX3} and \ce{PbX2} containing \ce{PbX6} octahedra, the anharmonic soft phonon modes with imaginary phonon energies were observed.
These indicate their structural instabilities, which can be associated with the distortion or tilting of \ce{PbX6} octahedra~\cite{Jong18jmca2,RYang17jpcl}.
For the case of perovskite solid solutions \ce{CsPbX3}, we find the BZ boundary (R, M, X) distortions, which are known to be antiferroelectric due to a cancellation of opposing polarizations induced in neighbour unit cells, and the zone centre $\gamma$-point instability which orginates a ferroelectric distortion~\cite{RYang17jpcl}.
It was accepted that these vibrational instabilities could be caused by the interaction between Cs and X atoms~\cite{Jong18jmca2} and stabilized by anharmonic processes at high temperature, providing an explanation of a series of phase transition in the perovskite crystals as varying temperature and pressure.
It is worth noting that as increasing the Br content $x$, the phonon energies increase in the overall branches without big anomaly.
However, the magnitude of phonon energy, especially of soft phonon mode, is not necessarily associated with the energetic barriers for the phase transition.

The phonon calculations were repeated with 11 different volumes for three kinds of solid solutions at each Br content $x$ so as to obtain the $F-V$ data within QHA and the Gibbs free energy $G(T, P)$ through fitting them to the 4th-order natural strain EOS.
Then, the Gibbs free energy of mixing was calculated as a function of Br content $x$ following Eq.~\ref{eq-miscib} and plotted as increasing temperature from 0 to 1000 K in Fig.~\ref{fig3}(d).
Under 300 K, the free energy differences were calculated to be positive in the whole range of Br contents, indicating that the solid solutions are not favourable to form but prefer to separate into their constituents \ce{CsPbI3} and \ce{CsPbBr3}.
As temperature increases, however, the vibrational contributions became enhanced and thus the solid solutions were found to be gradually stabilized with a negative free energy of mixing.
It should be emphasized that at room temperature, the solid solutions in all the Br contents were estimated to be stable from the phase separation, which is similar to other inorganic halide perovskite solid solutions~\cite{Jong18jmca2,Jung17cm}.

\begin{figure}[!t]
\centering
\includegraphics[clip=true,scale=0.5]{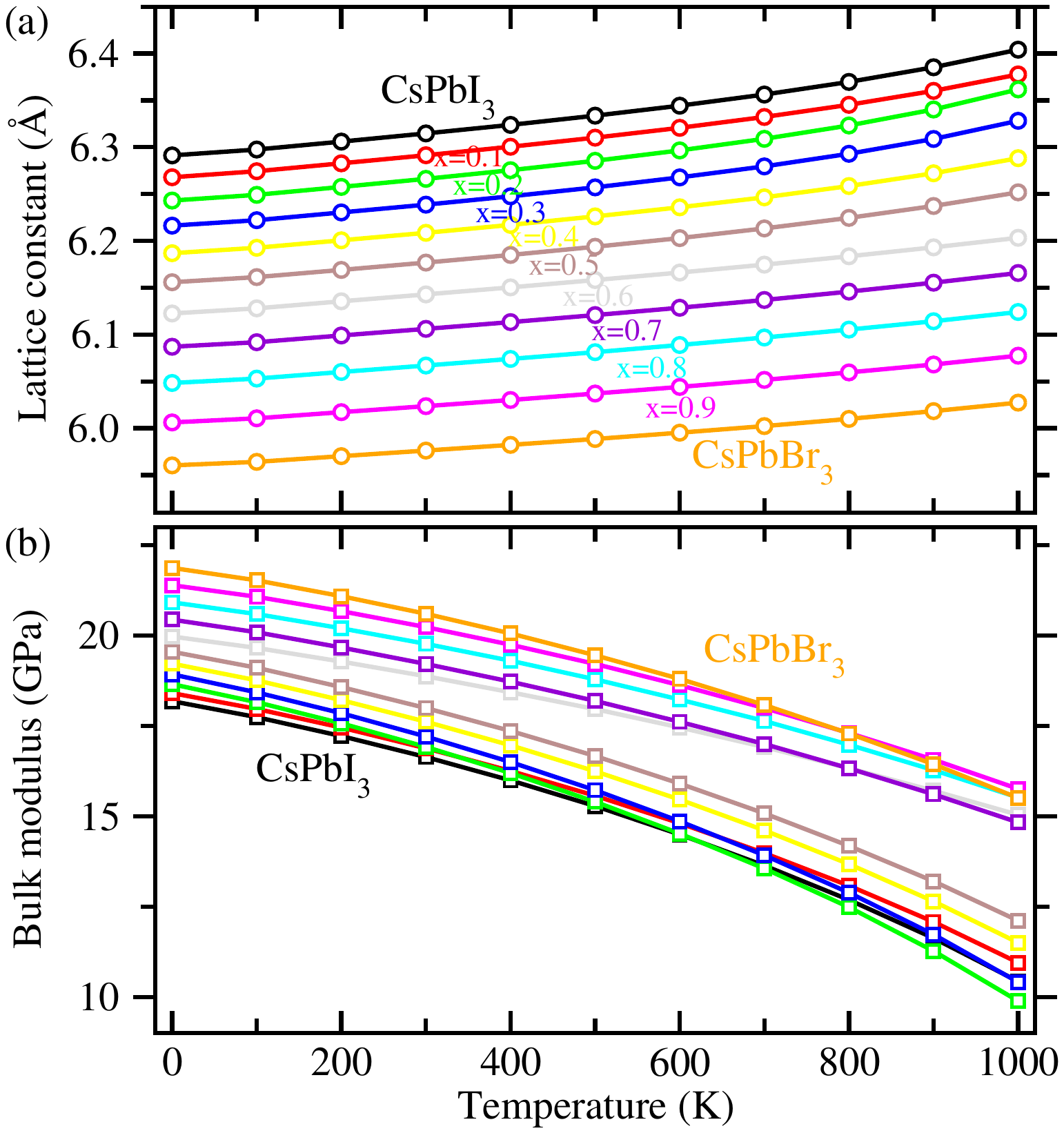}
\caption{(a) Lattice constants and (b) bulk moduli of solid solutions \ce{CsPbX3} (X = I$_{1-x}$Br$_x$) as increasing temperature, obtained by fitting the Helmholtz free energy versus volume data to the 4th-order natural strain equation of state.}
\label{fig4}
\end{figure}
Fitting the $F-V$ data to the EOS yields the lattice constants and bulk moduli dependent on temperature.
Figure~\ref{fig4} displays the lattice constants and bulk moduli of solid solutions \ce{CsPbX3} in cubic phase with various Br contents. As temperature increases, the cubic lattices were found to expand and the bulk moduli to reduce according to the nonlinear functions.
Such tendencies can be said to agree well with the general knowledge of the thermal expansion of solid materials.
It should be noted that as varying the Br content $x$, the increasing tendencies of lattice constants are systematic without any anomaly, whereas the decreasing trends of bulk moduli seem to be in disorder at higher temperatures possibly due to a numerical noise.

\subsection{Chemical stability upon decomposition}
\begin{figure*}[!th]
\centering
\begin{tabular}{cc}
\includegraphics[clip=true,scale=0.097]{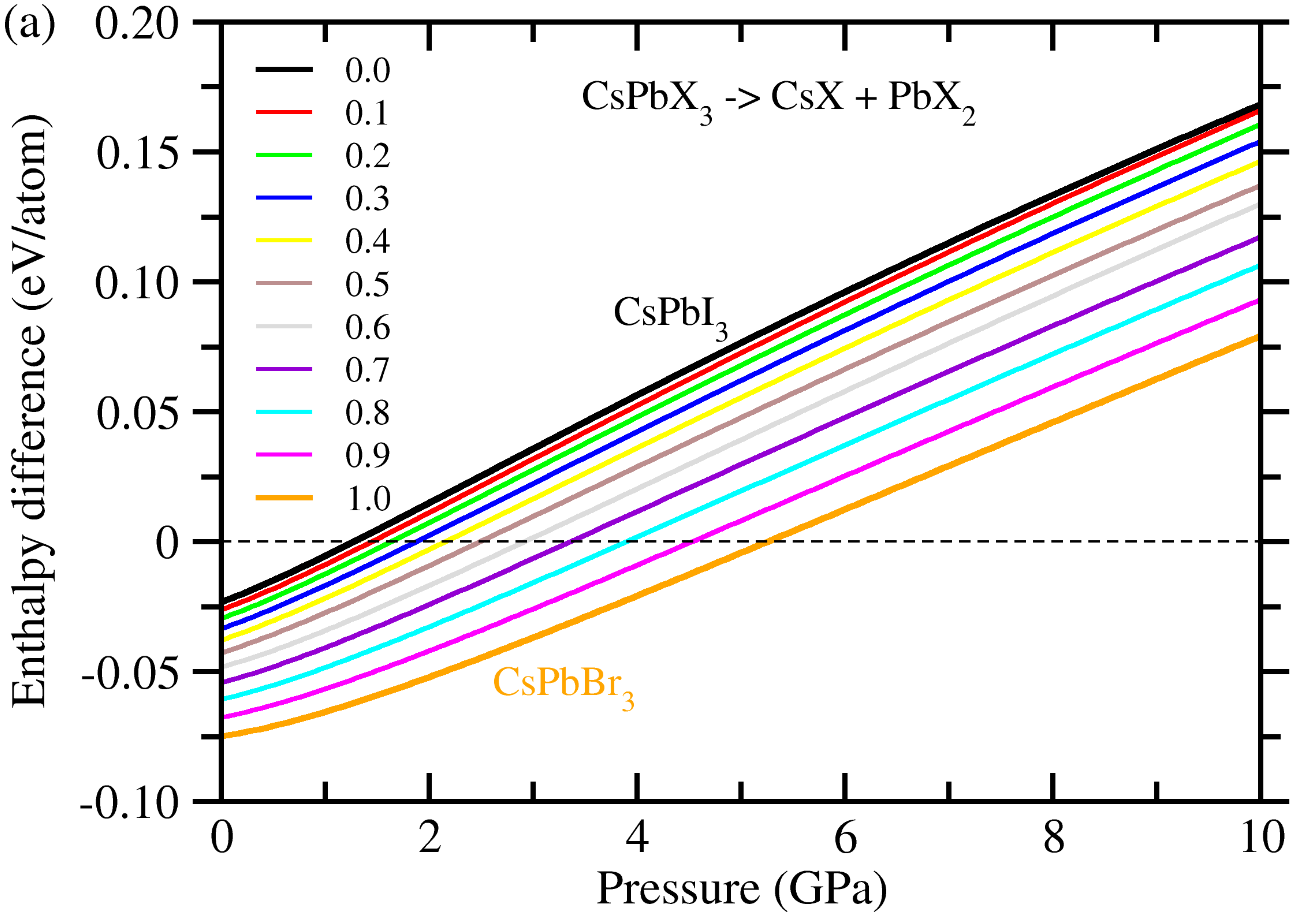} &
\includegraphics[clip=true,scale=0.149]{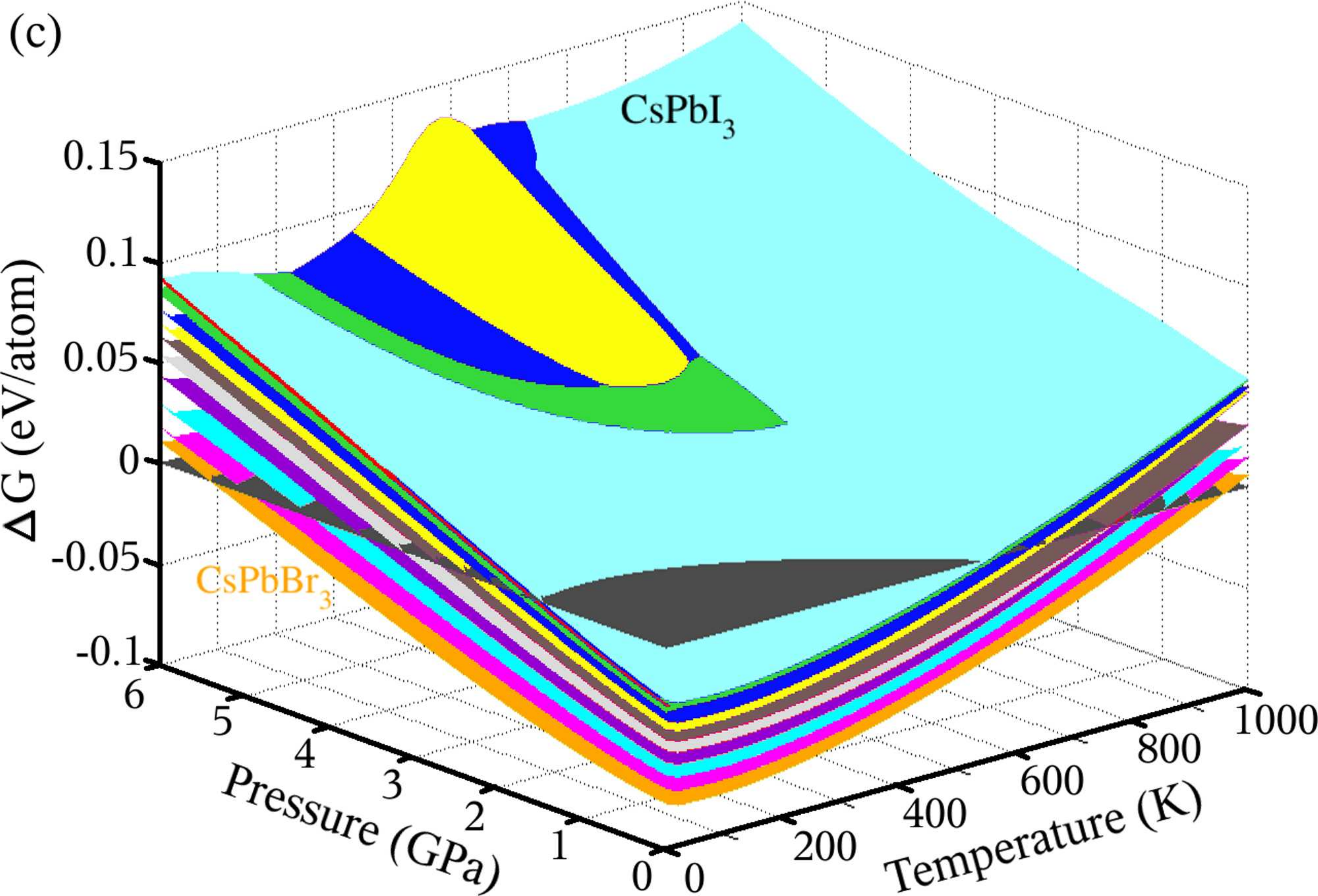} \\
~\includegraphics[clip=true,scale=0.097]{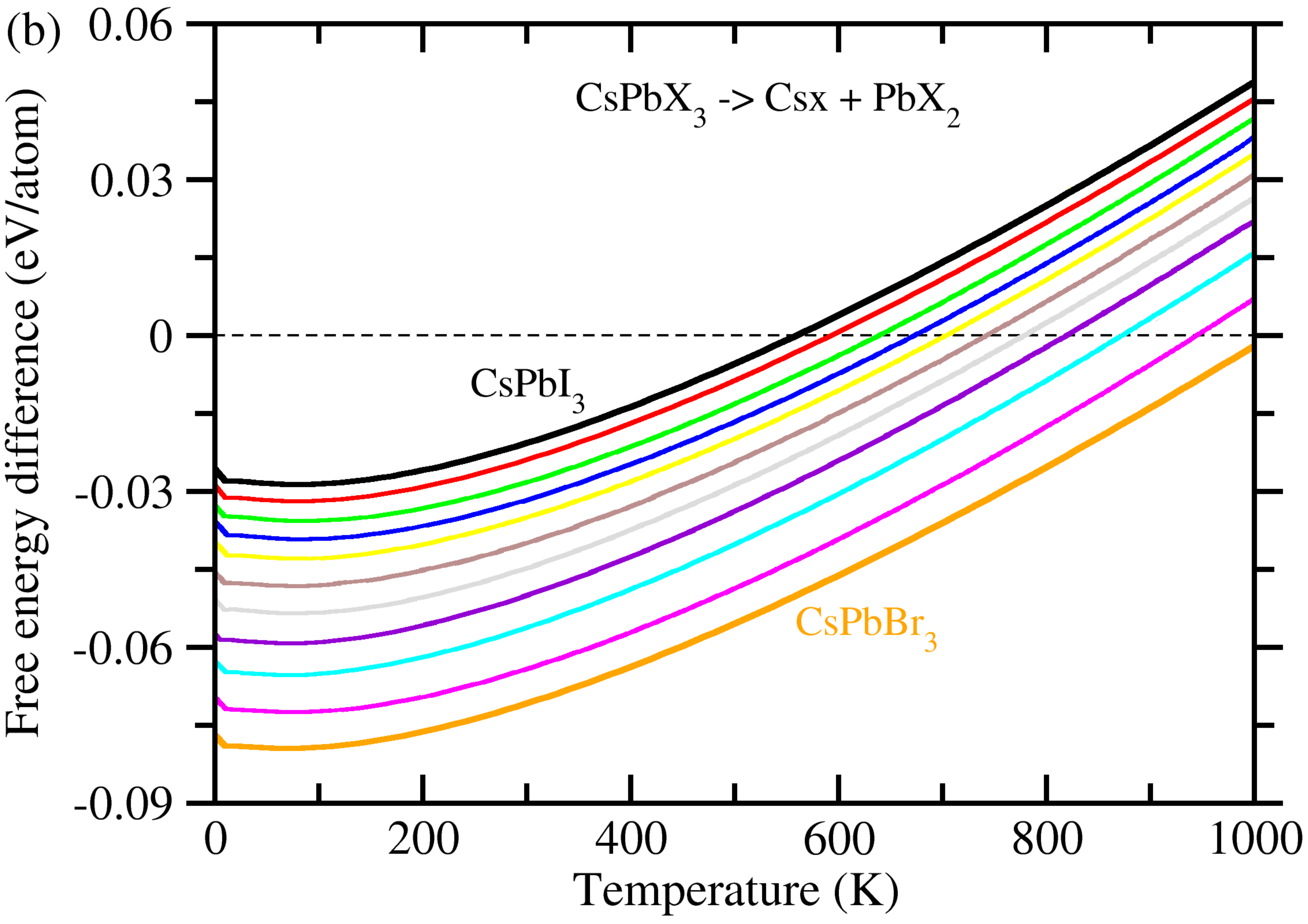} &
\includegraphics[clip=true,scale=0.129]{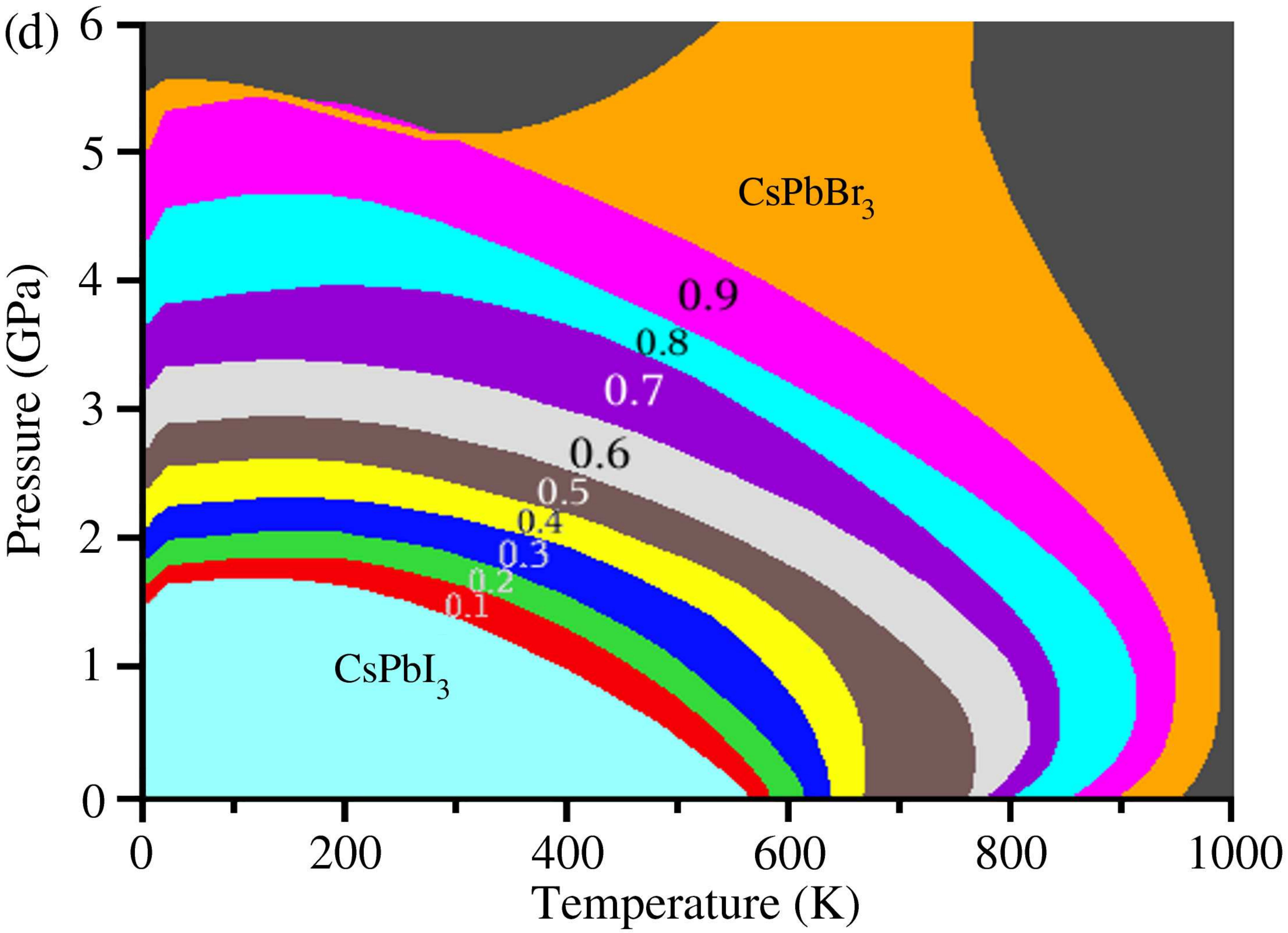}
\end{tabular}
\caption{(a) Enthalpy difference ($\Delta H$) as a function of pressure at zero temperature, (b) Helmholtz free energy difference ($\Delta F$) as a function of temperature at zero pressure, and (c) Gibbs free energy difference ($\Delta G$) as a function of temperature and pressure, for decomposition of \ce{CsPbX3} into its constituents CsX and \ce{PbX2} with increasing the Br content $x$ from 0 to 1. (d) Projection of Gibbs free energy difference onto the plane of $\Delta G=0$.}
\label{fig5}
\end{figure*}

In addition to the phase instability, the halide perovskites were also known to be readily decomposed into their constituent compounds under exposure to light, heat and moisture, representing the major obstacle to long-term outdoor application of PSCs.
The chemical stability can be estimated by calculating the difference of thermodynamic potentials such as internal energy, enthalpy, and free energy for decomposition of \ce{CsPbX3} into its constituents CsX and \ce{PbX2}.
When fixing temperature at 0 K, the enthalpy differences as a function of pressure should be calculated through fitting the $E-V$ data to the EOS.
Figure~\ref{fig5}(a) displays the enthalpy differences ($\Delta H$) in the pressure range of 0 $\sim$ 10 GPa as increasing the Br content $x$.
For all the Br contents, the enthalpy differences were found to increase from being negative to positive according to almost linear functions of pressure.
In the case of \ce{CsPbI3}, the turning point of $\Delta H=0$ was determined to be 1.28 GPa, indicating the pressure range of 0$-$1.28 GPa for being stable upon chemical decomposition.
As the Br content increases, the turning point was found to shift upward, arriving at 5.26 GPa for \ce{CsPbBr3}.
By regression, we obtained the quadratic function as $P_0(x)=1.318+0.629x+3.277x^2$ (GPa) rather than the linear function (see Fig. S6).
Therefore, it turns out that the stable pressure range upon chemical decomposition of solid solutions \ce{CsPbX3} widens quadratically as increasing the Br content at fixed temperature of 0 K.

When fixed pressure at 0 Pa, the Helmholtz free energies of solid solutions \ce{CsPbX3}, CsX and \ce{PbX2} with different Br contents were calculated by postprocessing their phonon dispersion curves and phonon density of states, and the differences ($\Delta F$) were determined as a function of temperature.
It should be noted that the imaginary phonon modes were not considered in the postprocess of phonon because they might not be associated with the chemical stability.
In Fig.~\ref{fig5}(b), we plot $\Delta F$ in the temperature range of 0 $\sim$ 1000 K for different Br contents.
Similarly to the case of $\Delta H$, the Helmholtz free energy differences were also found to increase but along the nonlinear functions of temperature rather than the linear functions.
For the case of \ce{CsPbI3}, the sign of $\Delta F$ changes from the negative to the positive at $\sim$560 K, indicating the stable temperature range of 0$-$560 K upon chemical decomposition in agreement with previous work~\cite{Jong18jmca2}.
As the Br content $x$ increases, the stable temperature point was found to increase as a quadratic function of $T_0(x)=571.82+241.25x+191.27x^2$ (K) by regression (see Fig. S6).
It is worth noting that the cubic \ce{CsPbBr3} crystal can be stable against chemical decomposition at high temperature over 1000 K.
When compared with the hybrid halide perovskites \ce{MAPbX3}, the stable temperature points could be said to be much higher, indicating that the PSCs made by using the inorganic halide perovskites \ce{CsPbX3} could have enhanced stability upon exposure to heat.

Finally, we investigated the effects of temperature and pressure synergetically on the chemical stability of \ce{CsPbX3} by calculating the Gibbs free energy differences ($\Delta G$) according to Eq.~\ref{eq-gibbs}.
In Fig.~\ref{fig5}(c), we show the 3-dimensional plot of $\Delta G(T, P)$ in the temperature range of 0 $\sim$ 1000 K and the pressure range of 0 $\sim$ 6 GPa for the solid solutions \ce{CsPbX3} with the increasing Br content $x$ from 0 to 1.
Figure~\ref{fig5}(d) displays the $P-T$ diagram obtained by projection of $\Delta G(T, P)$ onto the plane of $\Delta G=0$.
This 2-dimensional diagram contains the separate informations of chemical stability upon pressure ($P$-axis) and temperature ($T$-axis).
The cubic \ce{CsPbI3} crystal was found to be stable (that is, its decomposition into CsI and \ce{PbI2} could not occur) within the area of a quarter circle with the boundary points of $T=560$ K and $P=1.3$ GPa.
It was revealed that the area of chemical stability of \ce{CsPbX3} upon decomposition is gradually extended as increasing the Br content.
Therefore, it can be concluded that when mixing \ce{CsPbI3} with \ce{CsPbBr3}, the chemical stability of the resultant solid solutions \ce{CsPbX3} could be enhanced, with maintaining the photo-conversion efficiency as much as high.
We believe that such information will be useful to materials engineers (as a guide) who want to develop PSCs with enhanced efficiency and stability.

\section{Conclusions}
In this work we have performed the systematic investigation of structural, electronic, optical, vibrational properties and stabilities of inorganic halide perovskite solid solutions \ce{CsPb(I$_{1-x}$Br$_x$)3} as increasing the value of Br content $x$ with first-principles virtual crystal approximation calculations.
The calculated lattice constants of \ce{CsPbX3} were confirmed to agree well with the Vegard's law as they decrease along the linear function of Br content $a(x)=6.242-0.378x$ (\AA), and for the two end compounds, they were found to be in good agreement with the experimental values within the relative errors under 1\%, indicating a reliability of the constructed pseudopotentials of the virtual atoms.
With increasing the portion of exact exchange term from 0.01 to 0.33 in hybrid HSE functional, we obtained the band gaps varying according to the quadratic function of $E_g(x)=1.750+0.454x+0.180x^2$ (eV), which gives the values in good agreement with the available experimental data for $x=0, 0.22, 0.67, 1$ of Br content.
The frequency-dependent dielectric constants were determined by solving the Bethe-Salpeter equation or using the DFPT approach to obtain the light absorption coefficients and reflectivity, which show the systematic change as the Br content increases.
Through calculations of phonon dispersions, we analyzed their vibrational properties, revealing that \ce{CsPbX3} and \ce{PbX2} containing \ce{PbX6} octahedra exhibit the soft phonon mode and thus phase instability.
Finally the chemical stabilities were estimated by calculating the differences of thermodynamic potential functions such as enthalpy, Helmholtz free energy and Gibbs free energy, producing the $P-T$ diagram of being stable for \ce{CsPbX3} against the chemical decomposition.

\section*{Acknowledgments}
This work is supported as part of the basic research project ``Design of Innovative Functional Materials for Energy and Environmental Application'' (No. 2016-20) by the State Committee of Science and Technology, DPR Korea.
Computation was done on the HP Blade System C7000 (HP BL460c) that is owned by Faculty of Materials Science, Kim Il Sung University.

\bibliographystyle{elsarticle-num-names}
\bibliography{Reference}

\end{document}